\documentclass{emulateapj}

\newcommand{\etal}{{et al.}}
\newcommand{\eg}{{e.g.,}}
\newcommand{\ie}{{i.e.,}}
\newcommand{\tc}{3C\,48}
\newcommand{\tca}{3C\,48A}
\newcommand{\oiii}{[\ion{O}{3}]}
\newcommand{\kms}{km s$^{-1}$}

\begin{document}

\title{The Nature of Optical Features in the Inner Region of the 3C\,48 Host Galaxy\footnotemark[1]}

\footnotetext[1]{Based in part on observations made with the NASA/ESA Hubble
Space Telescope, obtained at the Space Telescope
Science Institute, which is operated by the Association of Universities
for Research in Astronomy, Inc., under NASA contract NAS 5-26555. These
observations are associated with program \# GO-09365. 
Also based in part on observations obtained at the Gemini Observatory, which is operated by the
Association of Universities for Research in Astronomy, Inc., under a cooperative agreement
with the NSF on behalf of the Gemini partnership: the National Science Foundation (United
States), the Particle Physics and Astronomy Research Council (United Kingdom), the
National Research Council (Canada), CONICYT (Chile), the Australian Research Council
(Australia), CNPq (Brazil) and CONICET (Argentina).}

\author{Alan Stockton\altaffilmark{2,3}, Gabriela Canalizo\altaffilmark{4}, Hai Fu\altaffilmark{2}, and William Keel\altaffilmark{5}}
\altaffiltext{2}{Institute for Astronomy, University of Hawaii}
\altaffiltext{3}{Cerro Tololo Inter-American Observatory}
\altaffiltext{4}{Department of Physics and Astronomy and Institute of Geophysics and Planetary Physics, University of California at Riverside}
\altaffiltext{5}{Department of Physics and Astronomy, University of Alabama}

\begin{abstract}
The well-known quasar \tc\ is the most powerful compact steep-spectrum radio-loud QSO
at low redshifts. It also has two unusual optical features within the radius of the radio jet
($\sim1\arcsec$):  (1) an anomalous, high-velocity narrow-line component,
having several times as much flux as does the narrow-line component 
coinciding with the broad-line redshift; and (2) a bright continuum peak (\tca) $\sim1\arcsec$
northeast of the quasar.  Both of these optical features have been conjectured to
be related to the radio jet. Here we explore these suggestions.

We have obtained Gemini North GMOS
integral-field-unit (IFU) spectroscopy of the central region around
\tc.  We use the unique features of the IFU data to remove unresolved emission at the
position of the quasar.  The resolved emission at the wavelength of the high-velocity
component is peaked $\lesssim0\farcs25$ north of the quasar, at virtually the same
position angle as the base of the radio jet. These observations appear to confirm
that this high-velocity gas is connected with the radio jet. However, most of the emission 
comes from a region where the jet is still well collimated, rather than from the regions
where the radio maps indicate strong interaction with an external medium.

We also present the results of
{\it HST} STIS spectroscopy of \tca. We show that  \tca\ is 
dominated by stars with a luminosity-weighted age of $\sim1.4\times10^8$ years, 
substantially older than any reasonable estimate
for the age of the radio source. Our IFU data indicate a similar age.
Thus, \tca\ almost certainly cannot be attributed to
jet-induced star formation.  The host galaxy of \tc\ is clearly the result of a merger, 
and \tca\  seems much more likely to be the
distorted nucleus of the merging partner, in which star formation was
induced during the previous close passage.
\end{abstract}





\section{Introduction}

\tc\ ($z=0.369$) is a remarkable quasar quite aside from the fact that it was the first
to be identified \citep{mat61,mat63}.
As seems to be the case with a number of  ``prototypes,'' it is far from typical. 
The original identification of \tc\ depended in part on
the small size of its radio source, and \tc\ and 3C\,277.1 remain the only examples of 
compact steep-spectrum (CSS) radio sources among powerful quasars at redshifts
$<0.5$. The host galaxy of \tc\ is unusually large and bright in comparison
with those of other low-redshift quasars \citep{kri73}. It is now recognized that
these host galaxy properties are largely a result of 
a major merger \citep{sto87,can00,sch04}.  \tc\ was the first quasar for
which an extended distribution of ionized gas was observed \citep{wam75}, 
and it still is one of the most luminous examples of
extended emission among low-redshift QSOs \citep{sto87}.
It is one of a handful of previously identified QSOs to have been shown
by IRAS also to be ultraluminous IR galaxies.  Star formation is currently
underway at a prodigious rate in the inner part of the host galaxy 
\citep{can00}, and there are still massive reserves of
molecular gas \citep{sco93,win97}.  All things taken
together indicate that we are witnessing a brief but important phase in the
history of a quasar, a time when host-galaxy star formation is near its peak,
the radio jet is breaking through the dense material in the inner part of the
host galaxy, and UV radiation from the central continuum source has just
recently become visible along many lines of sight.

\citet{wam75} noticed that the [\ion{O}{3}] profile in \tc\ was significantly blueshifted relative to
the broad H$\beta$ line.  Higher-resolution spectroscopy of the quasar
\citep*{cha99,can00}  shows the reason for this apparent shift:
in addition to the usual nuclear narrow-line region, there is very luminous extended
high-velocity gas within about 0\farcs5 of the nucleus.  \citet{cha99} presented evidence
from their integral-field-unit spectroscopy that the high-velocity gas has been 
accelerated by the interaction of the radio jet with the surrounding medium.
They discussed the variation of relative line ratios, velocities, and line widths
of the high-velocity and systemic-velocity components. \cite{can00} found that the high-velocity
gas was resolved on a slit through the nucleus at position angle $-28\arcdeg$, 
with an extent of $\sim0\farcs5$ and a discernible velocity
gradient.  But neither of these studies had sufficient spatial resolution to explore
the detailed relation between the radio jet and the high velocity gas.

\citet{bor82,bor84}
long ago showed the presence of strong Balmer absorption in the \tc\
host galaxy.  \citet{can00} examined the host galaxy of \tc\ in more detail
with the Keck Low-Resolution Imaging Spectrograph (LRIS;
\citealt{oke95}) and showed that the
spectrum of the stellar population had a classic ``K+A'' character. 
The fact that they saw
Balmer absorption lines from a relatively young stellar population
as well as evidence for an older population (\eg\ from the
Mg\,I$b$ feature at $\sim5180$ \AA) meant that
they could do a decomposition of the two components (using stellar
synthesis models, such as those of \citealt{bru93,bru03}) and estimate,
at least in a relative sense, the time that had elapsed since the end of the major starburst phase.
With sufficient signal-to-noise, it is possible to carry out such a decomposition
at each point in the host galaxy, building up a map of starburst ages
along lines of sight through the host galaxy
(strongly weighted towards the time at which the most recent major episode
of star formation ceased).  \citet{can00} were able to determine
starburst ages and stellar radial velocities in 32 distinct regions.

\begin{figure}[!t]
\epsscale{1.1}
\plotone{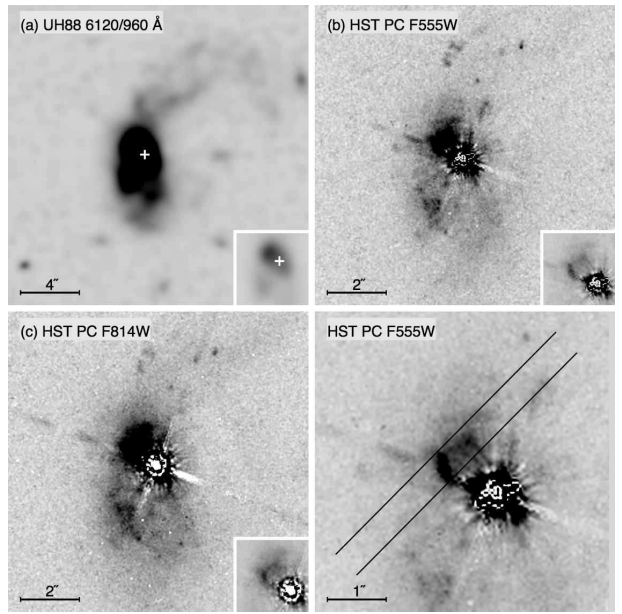}
\figcaption{{\it a---}Our deep line-free-continuum groundbased
image of \tc, after removing the quasar nucleus with the STSDAS procedure 
{\it cplucy}.  The cross
marks the position of the quasar.  \tca\ lies about 1\arcsec\ NE of the
nucleus (North is up and East to the left).
All insets show the central region of
the corresponding main panel at lower contrast.
{\it b} and {\it c---}HST PC images from the HST archive.  The nucleus
has been removed with synthetic Tiny Tim PSFs, so there are large residuals
in the inner 0\farcs5 radius and in the diffraction spikes.  The scale of
these panels is enlarged by a factor of 2 from that of {\it a}.  The filter
bandpasses are indicated.  {\it d---}Enlargement (by another factor of 2)
of the inner region of the HST PC F555W image, showing the position and
width of the STIS slit.
\label{hstfield}}
\end{figure}

But one of the most interesting features in the host galaxy was too
close to the quasar to permit sufficiently low levels of
contamination from quasar light to extract the galaxy spectrum.  This is
the luminosity peak \tca\ $\sim1\arcsec$ northeast of the quasar,
first noticed by \citet{sto91} and confirmed by subsequent
groundbased and HST observations \citep{cha99,kir99,boy99,can00}.
One of our ground-based images is shown in Fig.~\ref{hstfield}$a$,
and {\it HST} PC images obtained from the archive are shown in
Fig.~\ref{hstfield}$b$ and $c$.
\citet{sto91} originally suggested that \tca\ was
the secondary nucleus from a merging companion; this suggestion was
consistent with the presence of the tidal tail to the northwest, as well as
evidence that we are seeing the \tc\ host galaxy near the stage of final
merger \citep{can00}. The attempt to model the merger by \citet{sch04} shows
that there is short period during which the two nuclei would have a separation
and position angle roughly similar to that observed for \tc\ and \tca.
On the other hand, the morphology of the CSS radio jet \citep{wil91}
and the presence of the high velocity emission-line components near the quasar
\citep{cha99,can00} both suggest that
there is a strong interaction between the radio jet and the ambient gas
in the general region of \tca.  
\tca\ is dominated by continuum radiation; \ie\ the offset peak is seen
on continuum images but not on narrow-band [O\,III] images \citep{sto91,can00}. 
However, there is no detailed correspondence
between the morphology of the jet and that of \tca\ (the jet proceeds nearly
northwards rather than northeasterly), so the optical emission from \tca\ 
cannot be due to synchrotron radiation.  It clearly is unlikely to
be nebular thermal continuum from an interaction of the jet and the gas, 
because we would expect fairly strong emission-line radiation at reasonable
temperatures. \citet{cha99} find colors for \tca\ consistent with stars
with ages similar to the likely age of the radio jet ($\lesssim10^7$ years).
The HST images show that \tca\ is curiously distorted;
this distortion could result either from the shape of an overpressured
region resulting from shocks from the radio jet or from the tidal stretching of the nucleus
of a merging companion.

Here we use Gemini Multi-Object Spectrograph (GMOS) integral-field-unit (IFU) 
spectroscopy to analyze more precisely the position and extent
of the high-velocity emission, and spectroscopy with the Space Telescope 
Imaging Spectrograph (STIS) on the {\it Hubble Space Telescope} ({\it HST}) 
to attempt to determine the nature of \tca. Our principal goals for the STIS 
observations are (1) to determine whether the
continuum radiation associated with \tca\ is indeed due to stars;
if so, (2) to determine the ages of the stellar populations involved
and whether there is an age discontinuity; and
(3) to measure velocities of the stellar component:  do these show
any evidence for distinct kinematical signature, such as might be expected
if \tca\ is the distorted nucleus of a merging companion.  We assume a flat
cosmology with $\Omega_m = 0.3$ and $H_0 = 70$ km s$^{-1}$ Mpc$^{-1}$.

\section{Observations and Data Reduction}

\subsection{Gemini North GMOS Integral-Field-Unit Spectroscopy}
We obtained two-dimensional spectroscopy of the central region of \tc\ with the 
integral-field mode of GMOS on the Gemini
North Telescope (program ID no.~GN-2003B-C-5).  
We used the half-field mode with the B600 grating, which gave a 
field of $3\farcs3\times4\farcs9$ and a wavelength range of 4318--7194 \AA.
The lenslets have a width of 0\farcs2, and the CCD pixels correspond to 0\farcs05.
The total exposure was 9000 s, obtained as 5 1800 s exposures.
Observations of the spectrophotometric standard star G191B2B provided the flux calibration.

The raw spectral images were processed with the standard Gemini IRAF routines to
give the flux calibrated data cube.  The 6256 spatial planes of the data cube have dimensions
of $66\times98$ pixels (0\farcs05 square) and are spaced at intervals of 0.46 \AA.
This data cube still suffers from the effects of
atmospheric dispersion.  Since the quasar was included within the field of
view, we were able to determine its centroid for each spatial plane of the data cube.
We fitted a low-order cubic spline curve to the plot of each of the coordinates as a 
function of wavelength and used these to generate correction offsets for each plane
of the data cube.  Applying these corrections generated a new data cube for which
all of the image planes were correctly registered.  The field of view for which the full
range of wavelengths is available is, of course, somewhat smaller than the nominal
field because of the atmospheric dispersion. This corrected data cube was used for
analysis when we had to deal with a large wavelength range. However, for our 
detailed analysis of the \oiii\ profile, we used the uncorrected data cube, as 
described in \S\ \ref{hivel}, in order to avoid unnecessary interpolation.  The final 
image resolution at 6800 \AA\ (near redshifted \oiii) was 0\farcs55 (FWHM).

\subsection{{\it HST} STIS Spectroscopy}

The STIS spectroscopy was carried out with the G430L grating, covering a nominal
region from 2900 \AA\ to 5700 \AA.  The slit was centered at a position 0\farcs97
east and 0\farcs50 north of the quasar, at a position angle of 44\fdg74.
The efficiency decreases quite drastically
towards the UV end of the spectrum, and the useful
spectral range for our project was restricted to 3630 \AA\ to 5685 \AA.  Because of the
low surface brightness of \tca, we used the 0\farcs5 slit and binned by 2 pixels in
the dispersion direction, obtaining a total of 21076~s of integration over 8 orbits.  The
spectra were dithered over 4 positions along the slit. The slit position, superposed on
the {\it HST} PC F555W image, is shown in Fig.~\ref{hstfield}$d$.  We also
obtained 626~s of integration on the quasar itself with the same configuration, but without
binning.

Each of the 8 individual spectra of \tca\ was first given a preliminary cleaning of cosmic
rays with {\sc lacos\_im} \citep{vDok01}. A few additional cosmic rays and other blemishes were
masked manually. Masks from both the automatic and manual cosmic-ray detection for each
frame were combined with dead- and hot-pixel masks generated from all of the frames.
The dithered spectra and the corresponding frame masks were then shifted to a common
position and coadded, using the IRAF task {\it imcombine} with {\it ccdclip} rejection
(in addition to rejection of masked pixels).  
The background, including an estimate of the stellar background
in the \tc\ host galaxy, was removed by subtracting a linear fit to regions on either side
of \tca. The background level amounted to roughly a third of the total extracted flux
in our spectral aperture, and we estimate that its scaling is accurate to $\sim20$\%.
The spectrum of the brightest part of \tca\ was extracted, using a region
6 pixels (0\farcs3) wide. Some of the remaining noise spikes were removed by fitting a 
160-piece cubic spline to the spectrum and rejecting points more than $2 \sigma$ away
from the fit. The redshift was roughly determined from the [\ion{O}{2}] $\lambda3727$ doublet,
and the spectrum was reduced to the approximate rest frame. This redshift was
corrected to a final value from fits to the stellar absorption-line spectrum.

\section{The Distribution of Extended Emission in the Nuclear Region}\label{hivel}

\begin{figure}
\epsscale{1.1}
\plotone{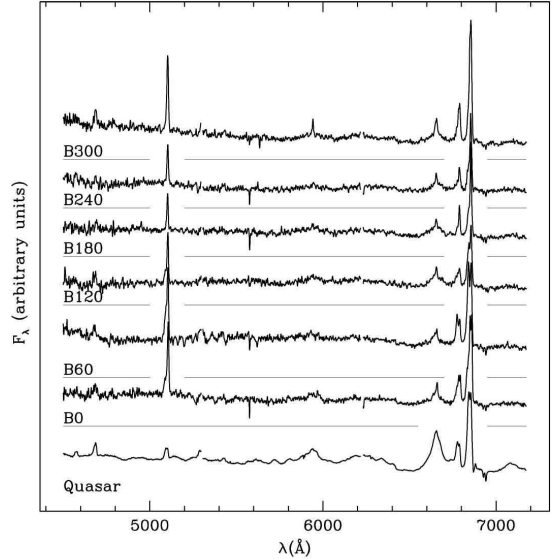}
\caption{Spectra extracted from the GMOS IFU datacube, as described in the text.
The quasar spectrum, shown at the bottom, has been scaled by a factor of 0.018 
to roughly match the other spectra, which are all on the same scale relative to each
other.  The spectra are shifted for clarity, with zero levels being indicated by the
horizontal lines. The off-nuclear spectra are all centered 1\farcs1 from the quasar,
at the position angles indicated.  No attempt has been made to correct for scattered 
light from the quasar.  The two slight gaps in each spectrum correspond to the
gaps between the CCD detectors, and the negative feature in some of the
spectra at 5577 \AA\ is due to imperfect subtraction of the strong airglow line
at that wavelength.\label{ifufull}}
\end{figure}

\begin{figure*}
\epsscale{1.0}
\plottwo{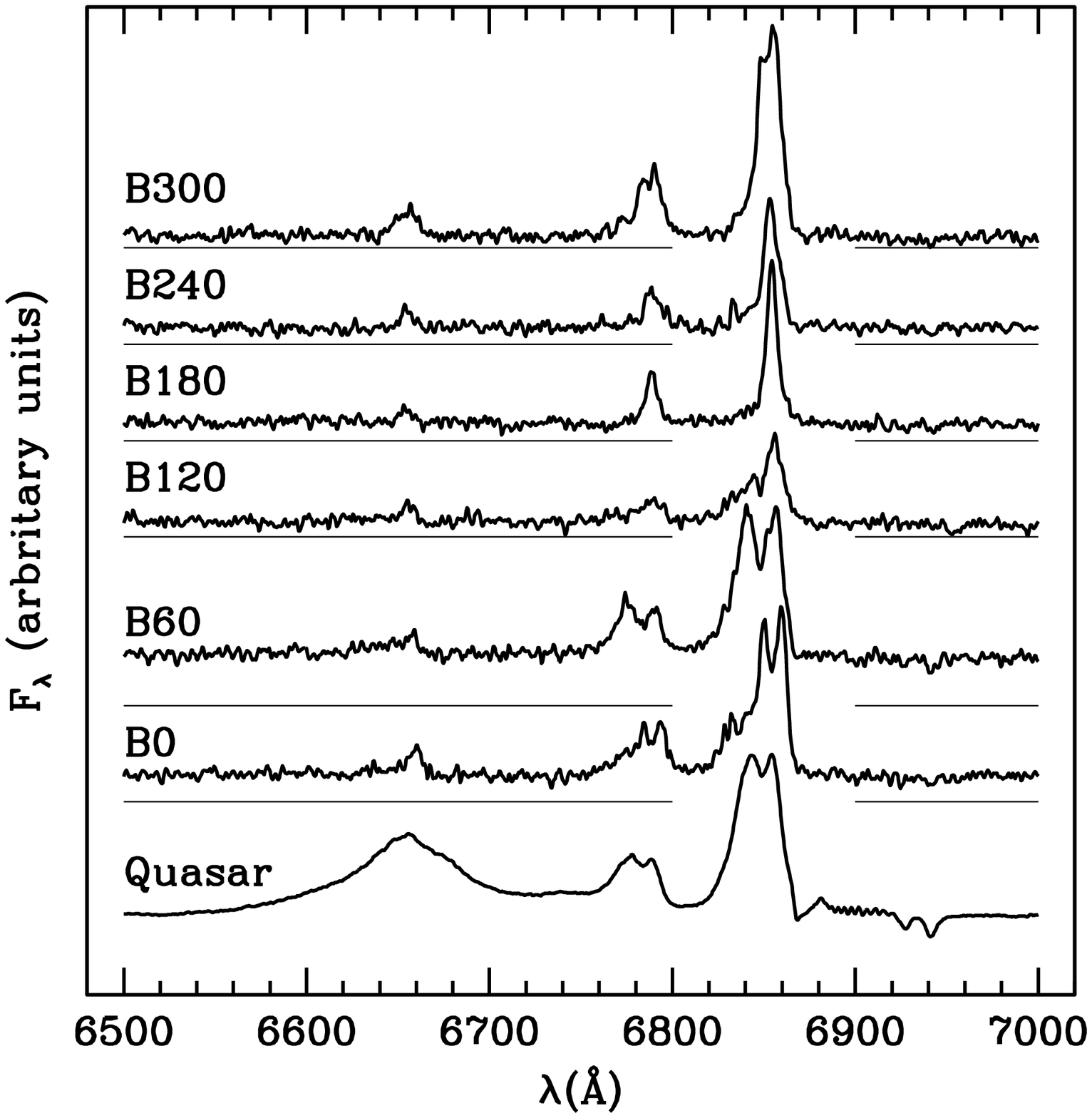}{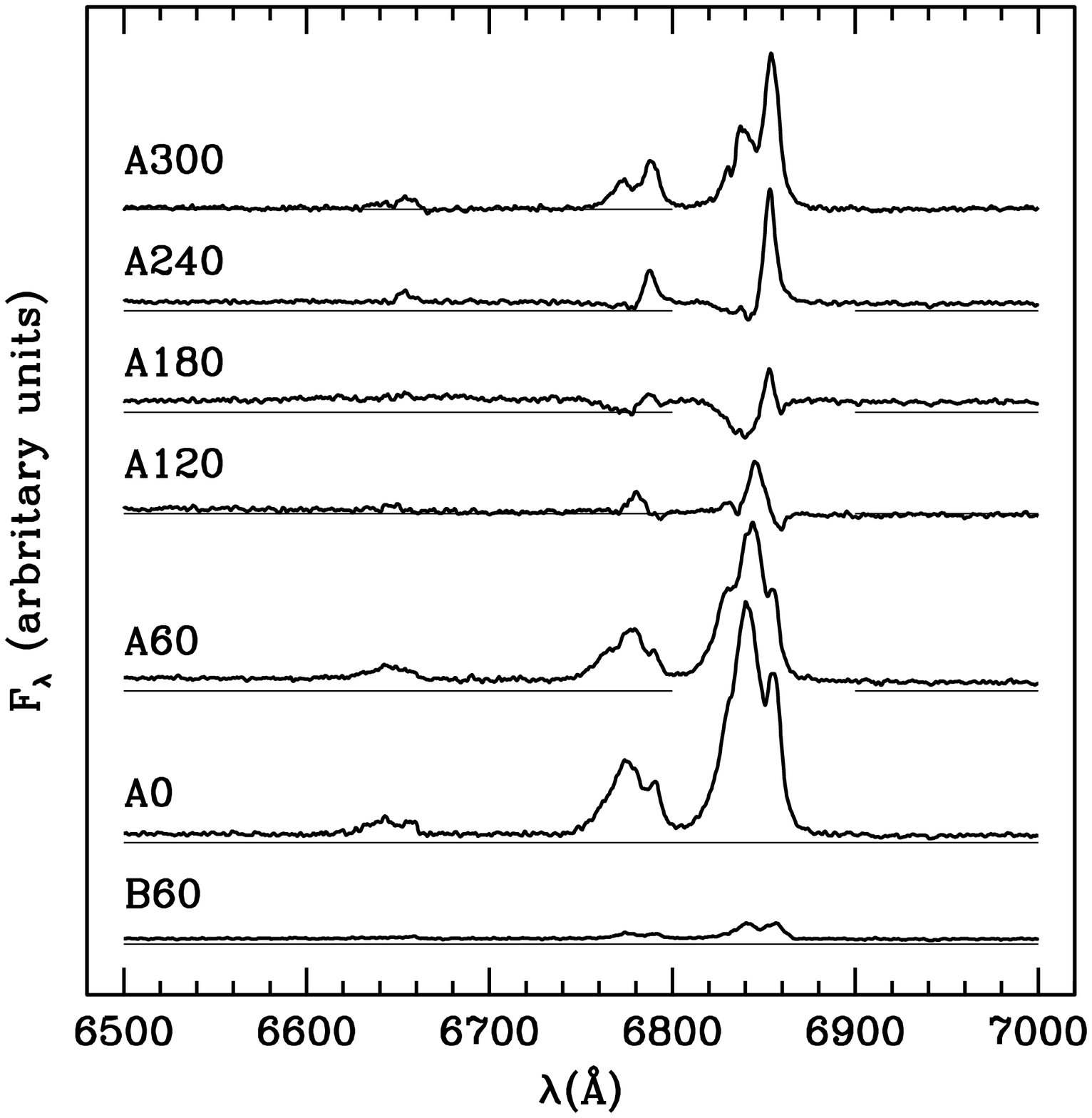}
\caption{({\it left panel}) Spectra of the region around the H$\beta$ and \oiii\ $\lambda\lambda4959$, 
5007 lines for the same apertures as are shown in Fig.~\ref{ifufull}; but, in this case,
a scaled quasar spectrum has been subtracted from each of the off-nuclear spectra
to remove the scattered quasar light, as judged by the broad H$\beta$ emission.
({\it right panel}) Similar quasar spectra for apertures at a radius of 0\farcs4 from the quasar.
The B60 spectrum at the bottom, shown at the same scale as the other spectra,
indicates the relative scaling for the two panels.
\label{ifuoiii}}
\end{figure*}

A full analysis of the emission over the field will be presented elsewhere; 
here, we concentrate only on emission very close to the quasar.  
For extracting spectra from our atmospheric-dispersion-corrected GMOS IFU
data cube, we define 13 apertures: one centered on the quasar, and two sets equally spaced
on circles around the quasar at position angles 0\arcdeg, 60\arcdeg, 120\arcdeg,
180\arcdeg, 240\arcdeg, and 300\arcdeg.  The inner circle has a radius of 0\farcs4,
and the outer circle has a radius of 1\farcs1; the inner apertures are labeled
A0, A60, A120..., and the outer apertures are labeled B0, B60..., where the
numerical part of the designation indicates the position angle of the aperture center from
the quasar.
All of the apertures were Gaussian weighted and were truncated at the FWHM 
of the Gaussian profiles, which was set to 0\farcs4.
We show the spectra of the quasar and the outer (B) ring of apertures in Fig.~\ref{ifufull}.
It was not possible to correct the off-nuclear spectra in Fig.~\ref{ifufull} for scattered 
light from the nucleus
because of the uncertain variation of the scattered contribution with wavelength.
However, we can make this correction (with typically about 20\% accuracy) for the
region in the immediate vicinity of H$\beta$ by subtracting a version of the quasar 
spectrum scaled to eliminate the broad H$\beta$ line in the residual.  We show the region
of H$\beta$ and \oiii\ $\lambda\lambda4959$, 5007 lines after this subtraction
for our 6 off-nuclear B apertures (centered 1\farcs1 from the quasar) in Fig.~\ref{ifuoiii} 
(left panel), as well as the same region for the quasar.  In the right panel of 
Fig.~\ref{ifuoiii}, we show similarly subtracted spectra for the A 
apertures (centered 0\farcs4 from the quasar).
For the apertures at 1\farcs1 radius (left panel in Fig.~\ref{ifuoiii}, the high-velocity 
component shows up most strongly in the B0 and B60
apertures; it is essentially completely absent in the B180, B240, and 
B300 apertures.  There is a clear velocity offset of the high-velocity component
between the B0 and B60 apertures, as well as evidence for more than
two velocity components in the former.  The high velocity emission is much stronger
in the inner apertures, particularly the A0 aperture, where the flux has over
10 times the highest value that it has in any of the outer apertures.  It is absent from the
A120, A180, and A240 apertures, but present in the
A60 and A300 apertures.  For this inner ring of apertures, the aperture
boundaries are adjacent to each other, so there will be unavoidable spill-over because
of the 0\farcs55 FWHM seeing disk.

At first sight, it may appear that is a problem with the subtraction procedure because of 
negative \oiii\ residuals, particularly in the A180 spectrum.  Note, however, that
these negative excursions are only at the wavelength of the high-velocity component.  The 
explanation is almost certainly that the quasar aperture encompasses high-velocity emission 
slightly extended to the north, which is then excess to the scattered nuclear quasar contribution to 
apertures on the south side.

In order to explore the spatial distribution of
the high-velocity  material in more detail, we analyze the image planes of the
data cube as we step across the \oiii\ line profile.  
Since we are here concentrating on a very
small range of wavelengths around the \oiii\ $\lambda5007$ emission complex, we can use the 
original data cube, uncorrected for atmospheric
dispersion, thus avoiding any smearing from subpixel interpolation.  The maximum shift due to
dispersion was confirmed to be less than 0.1 pixel ($=0\farcs005$) over this wavelength 
range. The general
approach is this:  we use the continuum on either side of the [\ion{O}{3}] $\lambda5007$ line
to precisely define the PSF of the quasar; we bin narrow wavelength intervals to form
images from the data cube as we step across the line profile, and we then subtract a scaled 
PSF from each of these to show the distribution of extended emission close to the quasar for
each part of the line profile.  

\begin{figure*}[!t]
\epsscale{1.0}
\plotone{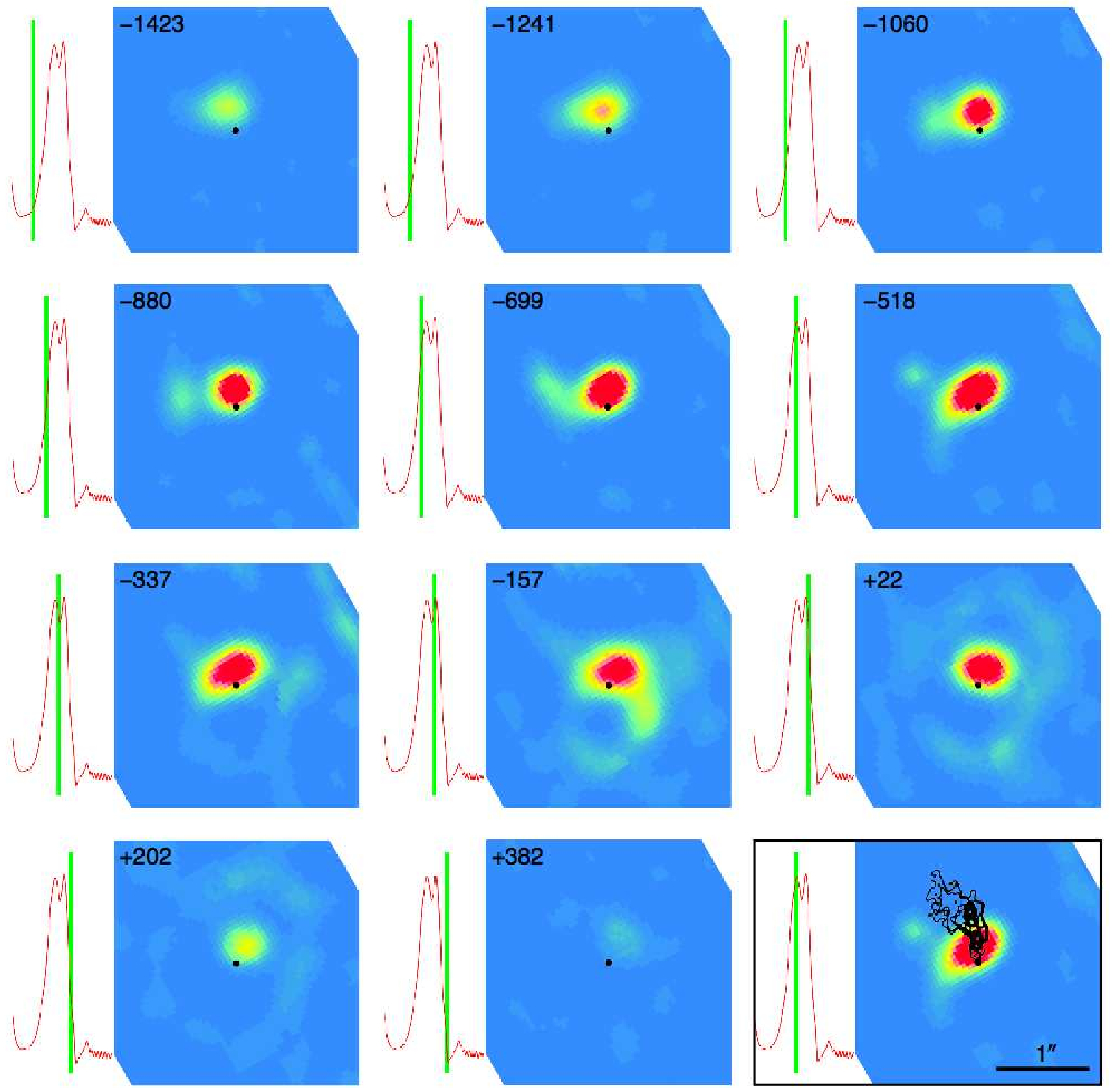}
\caption{The location of nuclear [\ion{O}{3}] extended emission in \tc. The unresolved
nuclear emission component has been removed from these images using {\sc plucy}
\citep{hook94}, where the PSF has been determined from the average continuum profile
on either side of the [\ion{O}{3}] line. The black dot marks the position of the quasar.
The [\ion{O}{3}] line profile is shown to the left of each
image, and the green bar indicates the wavelength region that has been summed to form
the corresponding image.  The number in the upper-left corner of the images gives the 
central velocity shift of the image, in km s$^{-1}$ in the quasar frame, from the systemic redshift of
0.36933.  The 5 GHz Merlin radio contours from \citet{feng05}
are shown superposed on the image near the peak of the high-velocity system
in the lower-right panel.  All images use a common intensity mapping and have
North up and East to the left.\label{nucemis}}
\end{figure*}

In more detail, we used narrow regions of continuum ($\sim20$ \AA\ wide)
immediately on either side of the emission line to define the PSF centroid at the line, 
but we used wider ($\sim100$ \AA) continuum regions somewhat farther away to obtain
better S/N for the PSF profile itself.  This high S/N PSF was then carefully registered
(at the sub-pixel level) to the PSF at the emission-line wavelength. We used 11 
4.6-\AA-wide slices to span the double line profile.  For each of these images, we ran
the two-component deconvolution task {\sc plucy} \citep{hook94}.  Briefly, {\sc plucy}
allows separate treatment of point sources (which can be removed) and a variable
background component, which is deconvolved via the Richardson-Lucy algorithm,
subject to an adjustable entropy constraint.  The result was that any unresolved nuclear emission
(\ie\ that having a distribution indistinguishable from that of the quasar continuum) was
removed, and only extended emission was retained in the residual images.
These images are shown in Fig.~\ref{nucemis}, along with the location in the
line profile each represents.  We also show the radio jet \citep{feng05} at the same scale.
What is immediately clear is that the high-velocity emission is concentrated to the north
of the quasar, precisely
in the initial direction of the radio jet, and mostly within the region where the jet appears
relatively unperturbed.  There is some slight variation in the position and shape
of the distribution of emission across the high-velocity line profile, and there is a tail of
emission at high velocities to the east, but the main emission peak remains closely aligned with
the radio jet. As the wavelength window shifts towards the systemic
redshift, the centroid of the extended emission moves more to the west, and more 
low-surface-brightness extended emission at greater distances becomes visible, including
a possible ring of emission at about 1\arcsec\ radius from the quasar.  Our decomposition of
the profile indicates that the narrow-line emission component at the systemic velocity is
dominated by the high-velocity component at all points in the composite profile.  The 
contribution of the systemic component is greatest (almost 50\%) near its peak, but
it drops off sharply at higher (as well as lower) velocities, because of the much broader
velocity width of the high-velocity emission.  We also 
expect the classical narrow-line region to be more closely centered on the nucleus and 
to be mostly unresolved.  Reconstruction of the line profile from the deconvolved, PSF-subtracted
images tends to bear out this expectation, although some residual systemic narrow-line emission still
remains, perhaps indicating that this region is slightly resolved (see also the profile in the A0
aperture in Fig.~\ref{ifuoiii}).
However, the shift in the centroid of the emission must still largely be due to a gradient in the
high-velocity component, rather than any extension of the classical narrow-line region.

From an aperture that includes essentially all of the high-velocity emission, our 
decomposition of the [\ion{O}{3}] profile gives a value of $491\pm40$ km s$^{-1}$
for the velocity offset between the two systems in the rest frame of the quasar.  
The [\ion{O}{3}] fluxes in the two components are $2.9\times10^{-13}$ and
$4.1\times10^{-14}$ erg s$^{-1}$ cm$^{-2}$, respectively, for the high-velocity and systemic
velocity components; \ie\ the high-velocity gas has $\sim7$ times the [\ion{O}{3}] luminosity
of that of the classical narrow-line region.

We have attempted to estimate contributions to the H$\beta$ profile from these two narrow-line
components.  We take the best-fitting Gaussian decomposition of the [\ion{O}{3}] profile, shift
these to the appropriate positions and widths for H$\beta$ at the same redshifts, and scale
the two components independently to obtain the maximum subtraction consistent with a smooth
and fairly symmetric core for the H$\beta$ residual.  This procedure gives a narrow-line 
H$\beta$/[\ion{O}{3}] ratio of $\sim0.1$ for the systemic-velocity component and
$\sim0.08$ for the high-velocity component.

We have also carried out decompositions of the [\ion{O}{2}] $\lambda3727$ doublet and the
[\ion{Ne}{5}] $\lambda3425$ line. The [\ion{O}{2}] case is complicated by the uncertainty in
the density-dependent doublet line ratio for each of the components, but the formal best fit
gives a line ratio at or near the low-density limit and a total flux $\sim16$\%\ of that of
[\ion{O}{3}] $\lambda5007$ for the high-velocity component and a line ratio near unity
(indicating an electron density of a few hundred per cm$^3$) and a total flux $\sim25$\%\
of that of [\ion{O}{3}] $\lambda5007$ for the component at the systemic velocity.  For the
[\ion{Ne}{5}] line, the high-velocity and systemic components have fluxes around 14\%\ and
20\%\ of those of their respective [\ion{O}{3}] profiles.  
Our ratios for the two velocity systems for H$\beta$/[\ion{O}{3}] and [\ion{O}{2}]/[\ion{O}{3}]
are quite consistent with those reported by \citet{cha99}. All in all, the ionization parameters of the 
two components seem roughly similar.

From our estimate of the H$\beta$ flux in the high-velocity component, and assuming an
electron density and temperature, the mass of the ionized gas in this component is given by
$$ 
M_{\rm H} = \frac{4\pi m_p f_{\rm H\beta} d_L^2}{\alpha_{\rm H\beta} n_e h\nu}{\rm ,} 
$$
where $m_p$ is the proton mass, $d_L$ is the luminosity distance,
$\alpha_{\rm H\beta}$ is the effective recombination coefficient for
H$\beta$,  and $h\nu$ is the energy of an H$\beta$ photon (Osterbrock
1989).  The recombination coefficient is roughly an inverse linear function of
temperature over the relevant temperature range; by assuming $T_e=10^4$ K, we are unlikely
to be in error from this parameter by more than a factor of two.

If we now assume that $n_e\approx100$ cm$^{-3}$ and use the H$\beta$ flux derived from
the H$\beta$/[\ion{O}{3}] ratio found above, we obtain $M_{\rm H}=7.4\times10^8$ 
$M_{\odot}$.  This is likely a fairly firm lower limit to the total mass.  Densities
much higher than 100 cm$^{-3}$ would be inconsistent with our fit to the [\ion{O}{2}] 
$\lambda3727$ profile.   
It remains quite possible, however, that most of the mass of the gas could have a density
$<10$ cm$^{-3}$, as it appears to be the case, for example, in the extended emission region 
around 4C\,37.43 \citep{sto02}. 
If this is true for the high-velocity gas associated with \tc\, its total mass could be as 
much as $10^{10}$ $M_{\odot}$.

Masses in this range can pose interesting constraints on acceleration mechanisms.
The high-velocity gas is centered at a projected distance $\lesssim1$ kpc from the
quasar with a differential radial velocity of $\sim500$ km s$^{-1}$.  \citet{wil91} argue that
the weakness of the central radio component in \tc\ means that the radio axis is likely
closer to the plane of the sky than to the line of sight.  In this case, the projected 
distance will be essentially the physical distance, and the radial velocity will be 
a lower limit to the space velocity.  Interestingly, \citet{gup05} find \ion{C}{4} $\lambda1549$
absorption at $z=0.3654$ in the spectrum of \tc, which corresponds to a radial velocity
difference in the quasar frame of 780 km s$^{-1}$.  Thus, it would be reasonable to
use $\sim800$ km s$^{-1}$ as a lower limit to the propagation speed of the radio
plasma through the ambient medium.  A plausible lower limit to the kinetic energy of the
high-velocity gas is then ${E}_K\gtrsim 6.4\times10^{57}M_9 v_{800}^2$ erg,
where $M_9$ is the mass of the high-velocity gas in units of $10^9$ solar masses,
and $v_{800}$ is the bulk average velocity of the gas, relative to the quasar, in units of
800 km s$^{-1}$.  \citet{wil91} estimate the total energy of the radio plasma as
$>10^{58}$ erg, so the kinetic energy of the gas may account for a significant fraction
of the original energy of the jet.  In this case, it is clear that the gas can exert a major
influence on the development of the jet, as \citet{wil91} and \citet{feng05} infer from the
morphology of the jet itself.  Nevertheless, it is significant that most of the high-velocity
gas is closer to the quasar than the region where the radio
jet appears to be disrupted.

\section{The Spectrum of 3C\,48A}
\begin{figure*}[!t]
\epsscale{1.0}
\plottwo{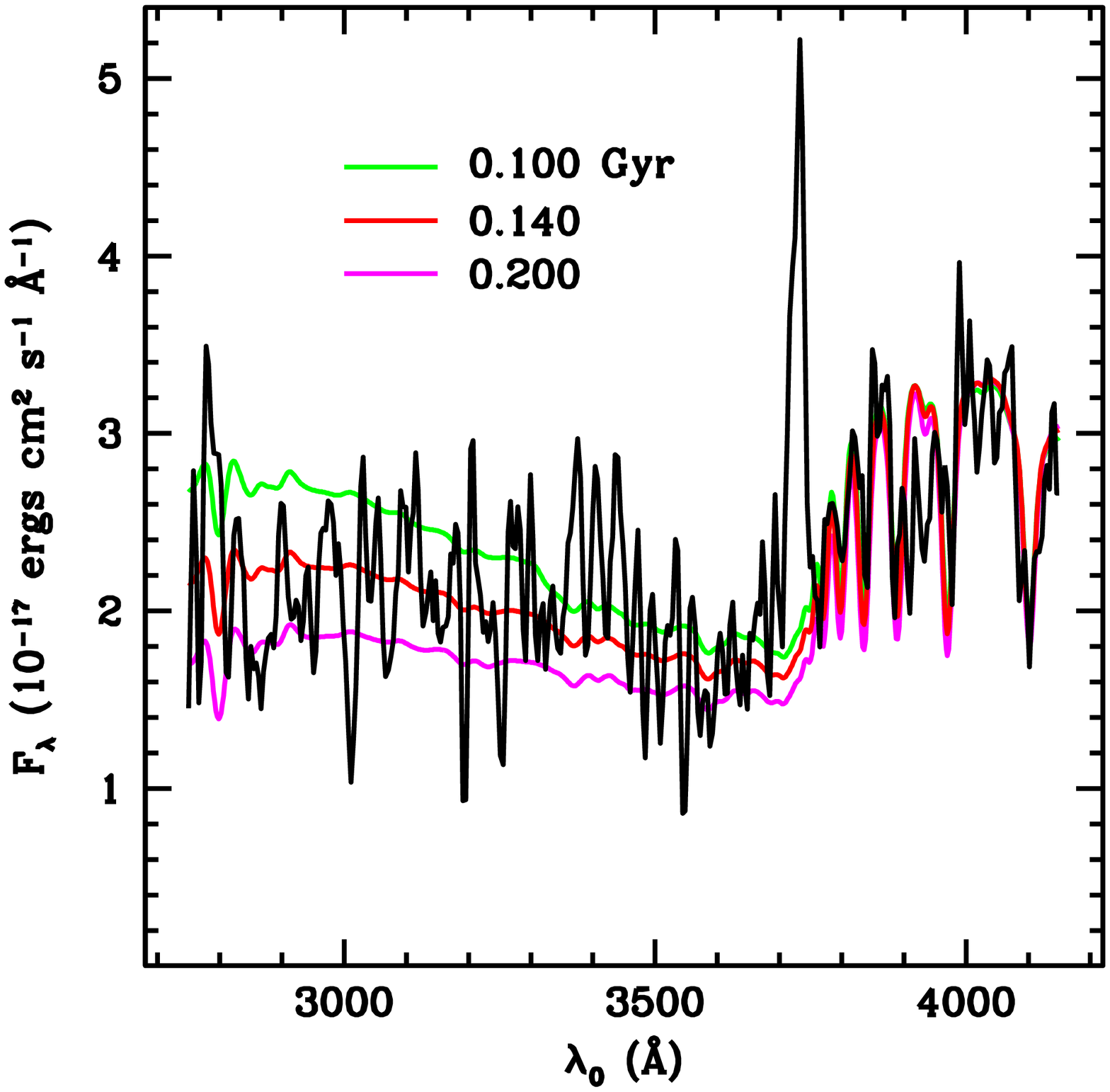}{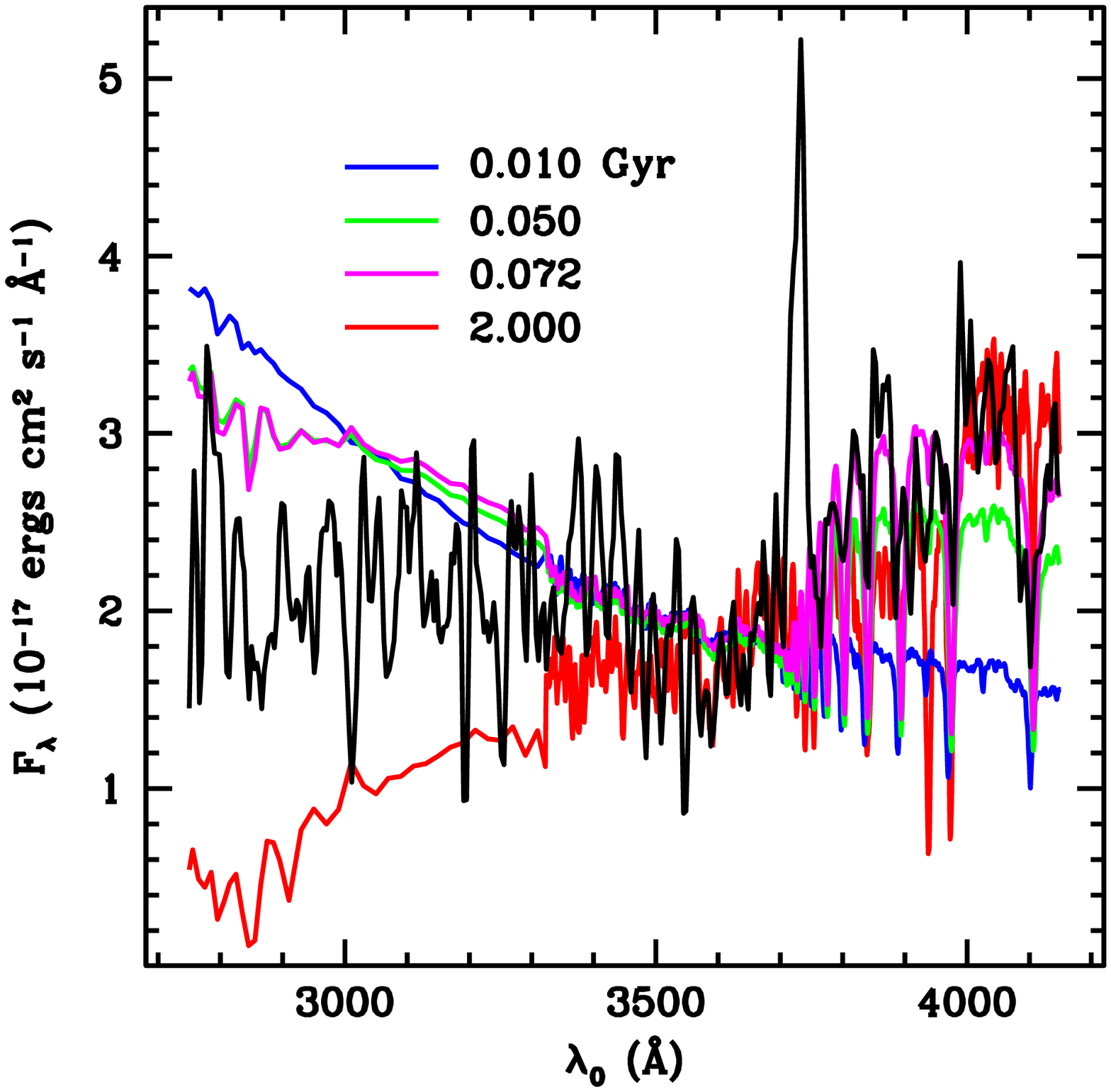}
\caption{Both panels show the {\it HST} STIS spectrum of \tca\ (black trace).  
The left-hand panel shows  a 140-Myr-old Bruzual \& Charlot (2003) instantaneous-burst
spectral synthesis model (red trace), and the green and blue traces show 100-Myr and
200-Myr models, respectively. The right-hand panel shows a range of more extreme models
with ages from 10-Myr (which would
be consistent with an estimate of the upper limit to the age of the stellar population in \tca\ 
obtained by \citealt{cha99}) to 2 Gyr.  Note that the Balmer lines and Balmer break constrain
the possibility of trying to fit younger populations with substantial reddening to the data.\label{stisspec}}
\end{figure*}

The final STIS spectrum of the high-surface-brightness arc-like structure in \tca\
is shown in Fig.~\ref{stisspec}. As had been indicated by a comparison of broad-band
and [\ion{O}{3}] {\it HST} WFPC imaging, the feature is dominated by continuum radiation,
although some [\ion{O}{2}] $\lambda3727$ emission is present.  A prominent 
H$\delta$ absorption line, seen at the long-wavelength end of the spectrum, together with
a minimum in the continuum near the Balmer limit,  suggests that the continuum is
largely due to moderately early-type stars.  Fits of spectral synthesis models confirm
this conclusion.  The absorption-line redshift, heavily dependent on H$\delta$, is 
$0.3693\pm0.0002$. In Fig.~\ref{stisspec} ({\it left panel}), we overplot on the 
spectrum \citet{bru03} solar-metallicity
instantaneous-burst models with ages ranging from 100 to 200 Myr, which appear
to bracket the luminosity-weighted average age of the stars (we also include a
10-Myr model for reference). The best fit, judged by the
slope of the continuum shortward of the Balmer limit, is about 140 Myr.
We have explored the possibility of younger stellar ages, combined with internal reddening,
as a means of trying to fit the observed spectrum. The fundamental difficulty with this approach
is the rapid decrease in the amplitude of the Balmer-limit break at ages below 100 Myr,
as shown in the right panel of Fig.~\ref{stisspec}. Because
of the short wavelength interval involved (from $\sim3600$ \AA\ to $\sim4000$ \AA), the break
cannot be significantly enhanced by reddening without doing unacceptable violence to the rest
of the spectrum, and this conclusion holds for any plausible reddening law.  It has previously been
noted (see, \eg\ \citealt{cha99}) that our particular line of sight to \tc\ must have little or no significant
reddening, in spite of the evidence for a major starburst in the host galaxy, and this conclusion
seems to hold for \tca\ as well, in spite of suggestions to the contrary by \citet{zut04}.

The fact that there is some evidence for a \ion{Ca}{2} K line stronger than in our models give
encouraged us to try combinations of an old stellar population with a younger one.  Such
a composite population might be expected to allow a younger age for the young population.
(Recall, however, that the host galaxy background near \tca\  has been subtracted off to
first order by our reduction procedure, so that any significant older population would likely 
have to be from \tca\ itself).
The general results of these experiments are that it is difficult to add any substantial fraction 
(by light) of old stars without unacceptable decreases in both the H$\delta$ strength and
the Balmer break amplitude, and, even with a fairly large fraction of old stars (20\%
at rest-frame 4050 \AA), the
younger component still could not be much younger than 10$^8$ years. 

\begin{figure}
\epsscale{1.0}
\plotone{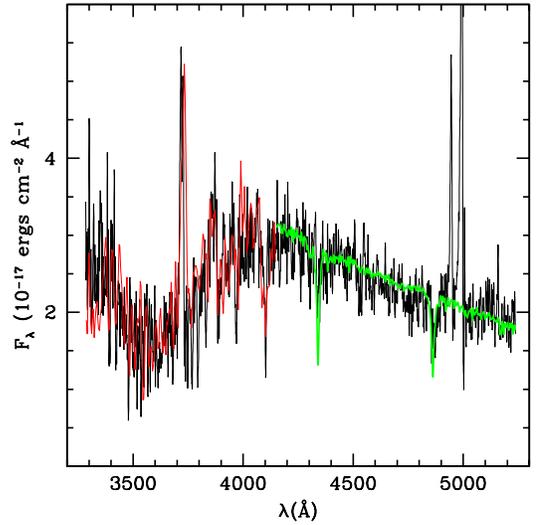}
\caption{Spectra and models of 3C\,48A.  The black trace is the spectrum of the
60\arcdeg\ aperture, after subtracting the mean of the other off-nuclear apertures
and scaling to match the {\it HST} STIS spectrum, which is shown in red.  The
green trace is the long-wavelength portion of the \citet{bru03} 140 Myr model, 
using the same scaling as the same model in the left panel of Fig.~\ref{stisspec}.
\label{ifua}}
\end{figure}

While the difficulties of attempting to obtain a clean spectrum of \tca\ from ground-based
observations are formidable because of scattered quasar light 
(and particularly the difficulty of correcting for its wavelength variation) as well
as contamination from other parts of the host galaxy, we were encouraged by the fact
that our 60\arcdeg\ aperture, which covers \tca, shows a substantially higher continuum 
level than the other apertures, even in the uncorrected spectra shown in Fig.~\ref{ifufull}.
We tried simply subtracting a straight mean of the other 5 off-nuclear apertures, which are all
at the same distance from the quasar.  If the quasar profile is closely symmetric, this procedure
should remove the scattered quasar light reasonably well.  It will also, to first order, subtract
the host-galaxy background, although there is certainly no guarantee that this will be very
accurate, because of variations in the host galaxy surface brightness.  The result is shown 
in Fig.~\ref{ifua}.

Naturally, one must be wary of putting too much trust in the details of this residual
spectrum because of variations among the 5 spectra that have been subtracted from it.
Nevertheless, it does agree with the general features of the STIS spectrum remarkably
well, and it reinforces the conclusion that the dominant stellar population in \tca\
must be old enough to show a substantial Balmer break.

These results are difficult to reconcile with the suggestion that \tca\ is in some way
connected with the radio jet.  The fact that \tca\ is dominated by stellar radiation
eliminates any possibility that it is due to any form of gaseous emission or emission
connected with relativistic electrons, such as synchrotron radiation or Compton upscattering
of photons. Dust scattering of starlight from elsewhere in the host galaxy is 
extremely unlikely, given that the most luminous source around by far is the quasar itself,
and no hint of scattered quasar radiation is seen in the STIS spectrum.  

A remaining possibility is jet-induced star formation, as has been suggested by
\citet{cha99}. This now also seems unlikely because of the age of the stars.  
An approximate dynamical time scale for the jet can be estimated from its length and the velocity of
the emission-line gas associated with it:
$\tau_{dyn}\approx10^6D_{kpc}v_{1000}^{-1}$ yr. For $D=5$ and $v=0.8$ (see \S\ \ref{hivel}), 
$\tau_{dyn}\approx6\times10^6$ yr. The apparent stellar age of \tca\ of $\gtrsim10^8$
years is thus much older than the likely age of the radio jet. It is also much older than the likely 
dynamical time indicated by the distance of \tca\ from the jet if it had been produced by it.
If the arc-like structure were to indicate that we are seeing the edge of a ``bubble'' of
star formation caused by the jet, the vector from the point of origin of the bubble to the
arc must likely be close to the plane of the sky. Because there is every indication that the
host galaxy of \tc\ is quite massive, yet the observed radial velocity range of the stars
is quite small \citep{can00}, stellar motions must be mostly in the plane of the sky.
In this case, transverse velocities of at least 100 km s$^{-1}$ must be expected for stars
in this feature at $\sim5$ kpc from the quasar. This means that over $10^8$ years, we
expect space motions of $\gtrsim10$ kpc, far larger than the $\sim3$ kpc observed 
projected separation between the end of the radio jet and \tca.

Furthermore, \citet{can00} had found a general trend of younger average stellar ages as
one moves inward in the \tc\ host galaxy, with stellar ages in the inner region of
$<10^7$ years, consistent with a stellar population dominated by a current
starburst. \tca\ therefore shows a clear age discontinuity with the stars in the host galaxy
just slightly farther out from the center.

There is suggestive, but not conclusive, evidence for a velocity discontinuity as well. 
The two host-galaxy regions with absorption-line velocities in the study by \citet{can00}
that are closest to \tca, C2 and B3 ($\sim2\arcsec$ northeast and east, respectively),
have stellar velocities $173\pm46$ and $170\pm54$ higher than that of \tca.  Velocity gradients
for the stars in central region of the \tc\ host galaxy appear to be quite low, so this
may be a significant difference.

All of this makes the possibility that \tca\ is actually the distorted nucleus of the
merging galaxy \citep{sto91} seem much more attractive. An age of $\sim10^8$ years
is not unreasonable for the time of the previous close passage prior to final merging. It
is also similar to the age of star-formation regions found in the leading edge of the
tidal tail ($\sim114$ Myr; \citealt{can00}), which might plausibly date from the same 
phase of the encounter.

\section{Comparison with Previous Investigations of 3C\,48}

The observations reported here are closely related to two previous spectroscopic studies 
of 3C\,48 and its host galaxy that have already been mentioned:  those of \citet{cha99} and
\citet{can00}. We have at various points noted comparisons of our present results with
these earlier studies, but it is worth discussing areas of agreement and disagreement
more systematically.  

\citet{cha99} obtained imaging spectroscopy of a $12\arcsec\times8\arcsec$
field around 3C\,48 with an IFU consisting of 655 optical fibers, each with a diameter
of 0\farcs4, mounted in the 
Multi-Object Spectrograph of the Canada-France-Hawaii telescope.  The total
exposure was 2700 s, and the seeing during the
observation was reported to be $\sim0\farcs6$.  The basic conclusions of this study were that:
(1) the broad, blueshifted [\ion{O}{3}] emission noticed in previous observations was
actually due to a combination of the normal narrow-line region at the systemic redshift and
a high-velocity component blueshifted by $\sim580$ \kms; (2) this high-velocity
component dominates the [\ion{O}{3}] profile in the center and over a region at least
0\farcs6 to the north and east of the quasar; (3) the location and energetics of this
high-velocity gas are consistent with its having been accelerated by the radio jet; (4)
the continuum colors of \tca\ indicate dominance by a stellar population younger than
$10^7$ years, with the star formation likely triggered by the radio jet; and (5) while much
of the emission-line gas appears to be photoionized by the central source, there is some
evidence for additional ionization from shocks.

\citet{can00} were mainly concerned with stellar population distributions and ages in the
host galaxy, but also discussed the emission lines present in their spectra.  They obtained
classical long-slit spectra with the Low-Resolution Imaging Spectrograph on the Keck II
telescope, with 6 different slit positions.  They also resolved the nuclear high-velocity
component, obtaining a velocity difference from the systemic narrow-line emission of
$563\pm40$ \kms.  While the limitations of using slits meant that they could not effectively map
its distribution, they were able to see that it had a strong velocity gradient over its
extent of $\sim0\farcs5$ in a slit through the nucleus at position angle 333\arcdeg.
In agreement with \citet{cha99}, \citet{can00} also felt that the general location and velocity
of this gas was strong evidence for a scenario in which it has been driven by the
radio jet.

\citet{can00} were not able to say anything about the stellar population of \tca\
because scattered light from the nucleus completely dominated the spectrum in
that region and could not be removed reliably, but they found quite young ages
($\sim4$ Myr, consistent with ongoing star formation) for regions in the host
galaxy they could measure that were just beyond \tca.  On the basis of the
colors reported by \citet{cha99} and on the morphologies of the {\it HST} images
of \tca, \citet{can00} stated, ``While it is still possible that this [\ie\ \tca] may be,
in fact, the distorted nuclear regions of the companion galaxy in the final stages
of merger, it now seems more likely that it, too, is related to the interaction of the
radio jet with the dense surrounding medium, as suggested by \citet{cha99}.''

The results we have given here support and refine the general conclusions of both of these
previous studies regarding the high-velocity gas.  With the higher resolution and
better S/N of our deep IFU observations, we have been able to map out the distribution of
the extended high-velocity gas in the region within $\sim1\arcsec$ of the nucleus and
at several points in its velocity profile.  These maps, shown in
Fig.~\ref{nucemis}, show that this emission peaks $\lesssim0\farcs25$ almost due north of
the quasar, near the base of the jet, and not, as might have been expected, near the point
of its obvious deflection and decollimation, roughly 0\farcs5 north of the quasar.

Our velocity difference between the high-velocity and systemic components of [{\ion{O}{3}}]
is $491\pm40$ km s$^{-1}$, compared 
with the previously reported values of $563\pm40$ km s$^{-1}$ \citep{can00} 
and $586\pm15$ km s$^{-1}$
\citep{cha99}.  These differences are at least partly due to different approaches to calculating
differential radial velocities from redshifts. \citet{can00} simply treated the redshifts of the
two components each as cosmological redshifts [for which $v/c={\rm ln}(1+z)$]
to find the difference in velocity, and
\citet{cha99} apparently did the same.  But since the velocity difference is actually due
bulk motion of material at essentially the same cosmological redshift as the quasar, it is 
more nearly correct to use the relativistic Doppler formula
$$
\frac{\Delta v}{c} = \frac{(1+z_1)^2 - 1}{(1+z_1)^2 + 1} - \frac{(1+z_2)^2 - 1}{(1+z_2)^2 + 1}
$$
to calculate the velocity difference.  If we use this approach for the previous determinations,
the \citet{can00} value becomes $519\pm40$ \kms, and the \citet{cha99} value becomes
$537\pm15$ \kms, both of which are now consistent with our new value. (We also
note that it is not clear how \citet{cha99} obtained their standard error of $\pm15$ \kms, since
their determination is dominated by their [\ion{O}{3}] $\lambda5007$ velocity difference, which
has a standard error of $\pm100$ \kms.  This small error may, in fact, have been a misprint, since, in
their ``Conclusions'' section, \citet{cha99} give the value as $\sim580\pm110$ \kms.)

\citet{cha99} had previously carried out a calculation of the total kinetic energy of the
high-velocity ionized gas and compared this value with the energy available from the
radio jet, similar to our calculation in \S\ \ref{hivel}.  The main difference in the calculations
is that \citet{cha99} assumed a mass for the gas given by \citet{fab87}, which happens to
agree well with the mass we obtain if the density in the emitting region is close to our
approximate upper limit of $n_e\approx100$ cm$^{-3}$ (this agreement is mostly fortuitous,
since \citealt{fab87} used different cosmological parameters and determined the density
simply by modeling the [\ion{O}{3}] to [\ion{O}{2}] emission-line ratio for a single-density 
medium). Our lower limit to the kinetic energy is about 20\% of the value ({\it not} intended
to be a lower limit) obtained by \citet{cha99}, so we are roughly consistent, in spite of
slightly different assumptions made in the respective calculations.  In both cases, the
minimum energy in the high-velocity gas comprises a significant fraction of the energy
of the radio jet, indicating that, if the radio jet is responsible for the current kinetic energy
of the gas, the gas is quite capable of affecting the dynamics of the jet.  However, especially
given the fact that most of the emission occurs in a region where the jet seems relatively
unperturbed, if the mass (and therefore the kinetic energy) of the gas should turn out to be
significantly larger than the lower limit, it could prove to be difficult to provide sufficient
energy from the radio jet alone.  If this were the case, one might be forced to appeal to 
some other form of
coupling the energy output of the quasar to the gas, the evacuation of a channel by the
radio jet merely providing a convenient path for the high-velocity gas to escape.  However,
at this stage, such speculations are premature.

On the question of the origin of \tca, we feel that our new estimate of the age
of the stellar population dominating the continuum light requires a re-evaluation of
the previous conclusions.  In our view, the balance of the evidence has swung back to the
original suggestion of \citet{sto91} that we are seeing the distorted nucleus of the merging 
companion, although with the modification that most of the light is coming from stars that
were likely formed from gas driven inwards during the first close pass of the interaction,
some $10^8$ years earlier.

Although we believe that our {\it HST} STIS spectrum clearly demonstrates the dominance
of stars with ages $\gtrsim10^8$ years, without more detailed information we can only
speculate on why \citet{cha99} found much younger ages from their estimate of colors for \tca. 
It seems possible that scattered light from the quasar may have had a more important
influence than they had estimated.  \citet{cha99} claimed that the scattered contribution to
\tca\ was $\sim20$\% at 5100 \AA\ and $\sim60$\% at 6900 \AA.  We have made
simple models of \tc\ and \tca\ using their quoted seeing of $\sim0\farcs6$.  With very
conservative assumptions (\eg\ a Gaussian rather than a Moffat profile for the quasar),
we find that the scattered contribution from the quasar would dominate that of \tca\ by
at least a factor of 2. A more realistic estimate would certainly be much larger. 
An additional possibility is that, with their rather large aperture, centered
1\farcs3 from the quasar, a bit beyond the center of light of \tca, at 1\farcs1,
and with ground-based seeing, the light \citet{cha99} measured may have been dominated
by the very young stellar populations in this region of the host galaxy \citep{can00}, rather
than light specifically from \tca. Experiments with the {\it HST} PC F555W image indicate that
over half of the light in such an aperture (apart from scattered QSO light) would have come
from host galaxy background stars rather than \tca.

\section{Summary and Discussion\label{sum}}

High velocity gas associated with high-redshift radio sources has often been attributed
to shocks from, or entrainment by, radio jets (\eg\ \citealt{sto96,sol01,lab05}; but see 
also \citealt{bau00}).  In most cases, the argument for these
mechanisms has been largely circumstantial: usually, there are essentially no other 
plausible sources for the observed velocities.  While hardly surprising, it is nevertheless
gratifying to find that our high-resolution imaging strongly confirms the positional relation between 
the high velocity gas and the steep-spectrum radio jet in \tc\ found by
\citet{cha99}.  Our main contribution here has been to show more clearly the distribution
of the gas as a function of velocity within the profile and to show that it appears to be
closely associated with the base and inner part of the jet, rather than the region farther
from the quasar, where high-resolution radio maps show the most obvious
deflection and decollimation.

In their two-dimensional simulations of the propagation of jets through inhomogeneous
media, \citet{sax05} show that it is quite plausible that CSS jets, such as that of 3C\,48,
do not require a direct collision with a dense cloud to suffer decollimation and deflection,
as has often been assumed.  
Instead, the jet, in overpressuring the intercloud medium, creates shocks that propagate 
into and reflect off of the dense clouds.  These reflected shocks can then redistribute
some of the forward momentum of the jet itself into transverse components. 

\citet{sax05} specifically consider the case of the \tc\ jet. They conclude that their model
of a jet progressing through a medium with a rather small filling factor of dense clouds
qualitatively matches the appearance of the \tc\ jet (modulo possible differences a fully
three-dimensional model might have from their two-dimensional simulations).  The jet 
is not completely disrupted; it roughly retains its initial direction, but it is deflected through
a small angle, and it becomes decollimated into a fan of material.

Taking this general picture as a working hypothesis, how does the high-velocity optical emission in
\tc\ fit in?  One possibility is that it is due to shock excitation of the gas in the dense clouds
responsible for decollimating the radio jet.  However, our conclusion
that  most of the high-velocity emission comes from a region quite close to the quasar,
before the jet shows significant disruption, means that the acceleration of the ionized
gas is likely not closely coupled with the deflection of the jet. 
In addition, it is not clear that the dense clouds themselves
will be accelerated to high velocities---whether they are or not depends on their size and mass distribution
and the momentum transfered to them by the bow shock of the expanding radio
cocoon.  Insofar as we have been able to
estimate the line ratios in the high-velocity component, they seem to be consistent with
photoionization by the quasar continuum, as well as with ionization by shocks (at least
with those for which a substantial fraction of the line radiation comes from photoionization
of pre-shock material by the thermal continuum of the shock itself).  An attractive option
is that most of the emission comes from photoionization by the quasar either of ablated material 
from small clouds or simply of ambient diffuse gas, which likely can be driven to high velocities 
by the wind generated by the expanding bubble around the jet.

The possibility that \tca\ might also be a consequence of the radio jet, in the form of
jet-induced star formation, now seems rather remote.  The age of the stars in this feature
is incompatible with jet-induced star formation by the current radio jet, and any appeal
to production by a hypothetical previous radio jet would have to explain why the feature 
remains so coherent and relatively close to the radio axis after $\sim10^8$ years.
Instead, a scenario in which we are seeing the distorted nucleus of the merging companion
galaxy seems much more likely. \citet{sch04} show that a plausible model of a merger of 
two equal-mass disk galaxies can both reproduce the approximate appearance of the
\tc\ host and, for a brief period, place the nuclei of the merging galaxies at the separation 
and approximate position angle of the quasar and \tca.  While it would be useful to
explore a larger range of parameter space to see if the model can be tuned to give a
more exact reproduction of the observed properties, this result is already sufficient to
give us confidence that this picture is basically correct. 

These results reinforce the conclusion that we are viewing \tc\ at a rather rare stage
in the life of a quasar and its host galaxy.  The age of the radio jet (which presumably
also dates the triggering of the current quasar activity) appears to be at most
a few $10^6$ years, and the two nuclei will fully merge on a time scale of $\sim10^7$
years.

\acknowledgments
Support for program \# GO-09365 was provided by NASA through a grant from the Space 
Telescope Science Institute, which is operated by the Association of Universities for 
Research in Astronomy, Inc., under NASA contract NAS 5-26555.
We thank the referee for a very careful reading of the paper and for making numerous
suggestions that materially improved the final version.
The authors recognize the very significant
cultural role that the summit of Mauna Kea has within the indigenous
Hawaiian community and are grateful to have had the opportunity to
conduct observations from it.

\clearpage

\end{document}